*Letter to the editor*

# Searching PubMed for articles relevant to clinical interpretation of rare human genetic variants

Andrew J. McMurry, PhD [1*]

[1] The Apache Software Foundation Dept. 9660 Los Angeles, CA 90084-9660 U.S.A.

*To whom correspondence should be addressed.
**Contact:** AndyMC@apache.org

***To the editor:***

While the speed and cost of genome sequencing has improved dramatically, the task of interpreting gene sequences for clinical purposes remains challenging [1-4]. Thousands of investigations into the pathogenicity of genetic variants have been completed and reported in peer-reviewed studies – however – which studies should be reviewed for each patient genome? The task of matching sequenced variants to the available evidence quickly exceeds human capacity.

Numerous challenges persist that delay clinical interpretation of human genetic variants, to name a few: (1) unstructured PubMed articles are the most abundant source of evidence, yet their variant annotations are difficult to query uniformly, (2) variants can be reported many different ways, for example as DNA sequence change or protein modification, (3) historical drift in annotations over time between various genome reference assemblies and transcript alignments, (4) no single laboratory has sufficient numbers of human samples, necessitating precompetitive efforts to share evidence for clinical interpretation.

Meaningful progress at the US National Library of Medicine and elsewhere provides improved capability to query and mine genetic databases[5-7] and PubMed articles[8-10] for rare genetic variants. Used together, these tools can be used to extract genetic variants from article text[10] and translate these mentions into formats recognized by the Human Genome Variation Society (HGVS) standards organization [11]. The implications of this capability are profound: PubMed articles can be indexed to provide near-real time lookup of articles relevant to a specific genetic variant.

**Search PubMed for articles relevant to a patient genome**

As an example, imagine a physician wants to rule out breast cancer for a patient, and orders genetic testing according to the approved guidelines for breast cancer[12]. In this example, a very rare genetic variant was found in the BRCA2 gene, with unknown disease pathogenicity. The standard list of widely cited databases is then checked: BIC – the breast cancer information core [13]; ClinVar – the NCBI repository of clinical variants [5]; and HGMD – the Human Gene Mutation Database[14]. It is rather typical that rare genetic sequence variants have no annotations available in any structured database, resulting in either a laborious review of potentially thousands of genetics articles or an uncertain interpretation -- Variant of Unknown Significance (VUS).

To aid clinical interpretation, Natural Language Processing (NLP) tools can be combined to extract mentions of genetic variants from PubMed article text [10]. Variant NLP tools exist for both rare and common genetic variants, germline and somatic. Relatively common genetic variants – variants having greater than 1% Minor Allele Frequency (MAF) -- may have standardized nomenclature and annotation available in dbSNP [7]. Intuitively, rare variants are rarely available in gene databases and thus present the greatest challenge for interpretation.

***From DNA to RNA to protein***

Results from published genetic studies may describe the variant of interest in terms of the genomic variant, DNA coding sequence, RNA transcript, or change in the resulting protein [15,16,17]. Modern variant tools such as tmVar [18] and SETH [9] extract mentions for all of these molecular types. These tools capture not only capture standard HGVS mentions of DNA, RNA, and protein mentions, but also improperly formatted mentions. Mappings between molecular types can be achieved using a wide range of open source tools. Importantly, variant mapping tools enable lookups of variants annotated in previous genomic references – older articles annotated using outdated genome and transcript assemblies can still be indexed and searched.

*Open source variant tools for searching PubMed*
Multiple Open Source packages can be combined[10] to provide as comprehensive a literature search as possible (Table 1). PubMed abstracts are routinely indexed by NCBI for gene names using GNorm+ [19,20] and textual mentions of genetic variants using NCBI tmVar [18]. NCBI conveniently provides downloadable files of the indexes in the NCBI public FTP site, which can be quickly downloaded and indexed in a local database [8]. For articles where abstract-text or full-text are available, SETH [9] can be used to extract variant mentions according to current and deprecated HGVS nomenclature.

**Table 1. Open access variant tools useful for searching PubMed**

| Software | Utility |
| --- | --- |
| PubTator | Download variants indexes from PubMed |
| Gnorm+ | Normalize gene name synonyms |
| SR4GN | Recognize species reference |
| tmVar | Extract variants from PubMed articles |
| SETH | Extract variants from PubMed articles |
| Variation Reporter | HGVS parsing and transcript mapping |
| Biocommons | HGVS parsing and transcript mapping |
| Mutalyzer | HGVS parsing and transcript mapping |